\documentstyle[12pt]{article}
\begin{document}
\tolerance=5000
\def\be{\begin{equation}}
\def\ee{\end{equation}}
\def\bea{\begin{eqnarray}}
\def\eea{\end{eqnarray}}
\def\nn{\nonumber \\}
\def\cF{{\cal F}}
\def\det{{\rm det\,}}
\def\Tr{{\rm Tr\,}}
\def\e{{\rm e}}
\def\etal{{\it et al.}}
\def\erp2{{\rm e}^{2\rho}}
\def\erm2{{\rm e}^{-2\rho}}
\def\er4{{\rm e}^{4\rho}}
\def\etal{{\it et al.}}

\  \hfill
\begin{minipage}{3.5cm}
OCHA-PP-159 \\
NDA-FP-75 \\
\end{minipage}

\vfill

\begin{center}
{\large\bf Holographic conformal anomaly with bulk scalars 
potential from d3 and d5 gauged supergravity}

\vfill

{\sc Shin'ichi NOJIRI}\footnote{nojiri@cc.nda.ac.jp},
{\sc Sergei D. ODINTSOV}$^{\spadesuit}$\footnote{
On leave from Tomsk Pedagogical University, 634041 Tomsk, RUSSIA. 
odintsov@ifug5.ugto.mx, odintsov@mail.tomsknet.ru}, \\
{\sc Sachiko OGUSHI}$^{\diamondsuit}$\footnote{
JSPS Research Fellow,
g9970503@edu.cc.ocha.ac.jp}
\\

\vfill

{\sl Department of Applied Physics \\
National Defence Academy,
Hashirimizu Yokosuka 239, JAPAN}

\vfill

{\sl $\spadesuit$
Instituto de Fisica de la Universidad de Guanajuato,
Lomas del Bosque 103, Apdo. Postal E-143, 
37150 Leon,Gto., MEXICO }

\vfill

{\sl $\diamondsuit$ Department of Physics,
Ochanomizu University \\
Otsuka, Bunkyou-ku Tokyo 112, JAPAN}

\vfill

{\bf ABSTRACT}

\end{center}

d3 and d5 maximally SUSY gauged supergravity is considered in the 
parametrization (flow) of full scalar coset where the kinetic term 
for scalars takes the standard field theory form and the bulk potential 
is an arbitrary one subject to consistent parametrization. From 
such SG duals we calculate d2 and d4 holographic conformal anomaly 
which depends on bulk scalars potential. AdS/CFT correspondence 
suggests that such SG side conformal anomaly should be identified 
with (non-perturbative) QFT conformal anomaly (taking account of 
radiative corrections) for the theory living on the boundary 
of AdS space. In the limit of constant bulk potential and single 
scalar, d4 result reproduces the known exact conformal anomaly 
corresponding to maximally SUSY super Yang-Mills theory 
coupled to ${\cal N}=4$ conformal supergravity. 

\newpage

d5 gauged supergravity which usually may be obtained as the truncation 
of d10 IIB SG is very useful in AdS/CFT correspondence\cite{AdS}. 
Classical solutions of such d5 theory describe RG flows in dual 
QFT living on the boundary of AdS bulk space. Many quantities of 
QFT dual may be found from SG side. In particular, one of the very 
important characteristics of boundary QFT is conformal anomaly. 
It can also be evaluated from SG side. Such calculation from 
d5 gauged SG with non-trivial single scalar and constant or 
non-constant bulk potential has been presented in refs.\cite{SN,LCA},
respectively. (Such version of d5 SG corresponds to special 
parametrization where scalars lie in one-dimensional submanifold 
of full scalar coset). In AdS/CFT set-up the scalars of gauged 
SG play the role of coupling constants or d4 scalars for boundary 
QFT. Hence, holographic conformal anomaly is very important as 
this gives presumbly the only way to get the exact conformal 
anomaly (with radiative corrections). Note in this respect that 
QFT Weyl anomaly with radiative corrections has been found so far 
for simple theories, like gauge theory with scalars, and only up 
to two- or three-loops. Comparison of such bulk and boundary side 
results may help in the explicit identification of dual QFT with 
the correspondent bulk configuration. From the other side, the 
holographic trace anomaly plays the role in the construction 
of local surface counterterms for gauged SG. Hence, it is also 
relevant in the evaluation of gravitational stress-tensor 
on the boundary.

In the present letter we evaluate d2 and d4 SG side conformal 
anomaly (where the bulk scalars potential is included) from 
d3 and d5 maximally SUSY gauged supergravity, respectively. 
Our formalism is general enough and may be applied to any 
gauged SG model. The only restriction is that we need free 
kinetic term for scalars, not the tensor dependent from scalars 
in front of $\partial_\mu\phi\partial^\mu\phi$ term, which is 
usual case in d5 gauged SG. That is why we consider the reductions 
from complete scalar space leading to such type kinetic term. As such 
type of reductions has been studied in ref.\cite{CGLP,MTR} we will limit
ourselves by the examples given here. 
Any other model of gauged SG with standard  (free) scalar kinetic term 
and arbitrary potential may be well studied in our formalism.

The dilatonic gravity action which may be considered 
as symmetric special flow of gauged supergravity where scalars lie 
in one-dimensional submanifold of full coset scalar space is given 
as follows: 
\be
\label{i}
S={1 \over 16\pi G}\int_{M_{d+1}} d^{d+1}x \sqrt{-\hat G}
\left\{ \hat R + X(\phi)(\hat\nabla\phi)^2
+ Y(\phi)\hat\Delta\phi
+ \Phi (\phi)+4 \lambda ^2 \right\} \ .
\ee
Here $M_{d+1}$ is a $d+1$ dimensional manifold (AdS$_{d+1}$ space ) 
whose boundary is a $d$ dimensional manifold $M_d$, 
where $d$-dimensional conformal field theory (CFT$_{d}$) lives 
and we choose $\Phi(0)=0$. As one can see the action contains 
an arbitrary dilatonic potential.

Let us choose the metric $\hat G_{\mu\nu}$ on $M_{d+1}$ and 
the metric $\hat g_{ij}$ on $M_d$ in the following form 
\be
\label{ib}
ds^2\equiv\hat G_{\mu\nu}dx^\mu dx^\nu
= {l^2 \over 4}\rho^{-2}d\rho d\rho + \sum_{i=1}^d
\hat g_{ij}dx^i dx^j \ ,\quad
\hat g_{ij}=\rho^{-1}g_{ij}\ .
\ee
Here $l$ is related with $\lambda^2$ 
by $4\lambda ^2 = d(d-1)/{l^{2}}$. 
If $g_{ij}=\eta_{ij}$, the boundary of AdS lies at $\rho=0$.
One should also add surface terms \cite{3}
to the above bulk action in order to have well-defined 
variational principle. However, these surface terms do not 
affect the calculation of Weyl anomaly via AdS/CFT 
correspondence \cite{AdS}, so we neglect these terms here.
  
We follow the same method of holographic anomaly calculation 
from SG side \cite{HS}. It has been generalized to the presence of 
a non-trivial dilaton in refs.\cite{SN,LCA}. Let us briefly describe 
this method below.
The action (\ref{i}) diverges in general since it
contains the infinite volume integration on $M_{d+1}$. 
The action is regularized by introducing the infrared cutoff
$\epsilon$ and replacing
\be
\label{vibc}
\int d^{d+1}x\rightarrow \int d^dx\int_\epsilon d\rho \ ,\ \
\int_{M_d} d^d x\Bigl(\cdots\Bigr)\rightarrow
\int d^d x\left.\Bigl(\cdots\Bigr)\right|_{\rho=\epsilon}\ .
\ee
One also expands $g_{ij}$ and $\phi$ with respect to $\rho$:
\be
\label{viib}
g_{ij}=g_{(0)ij}+\rho g_{(1)ij}+\rho^2 g_{(2)ij}+\cdots \ ,\quad
\phi=\phi_{(0)}+\rho \phi_{(1)}+\rho^2 \phi_{(2)}+\cdots \ .
\ee
Then the action is also expanded as a power series in $\rho$. 
The subtraction of the terms proportional to the inverse power of 
$\epsilon$ does not break the invariance under the scale 
transformation $\delta g_{ \mu\nu}=2\delta\sigma g_{ \mu\nu}$ and 
$\delta\epsilon=2\delta\sigma\epsilon$. When $d$ is even, however, 
the term proportional to $\ln\epsilon$ appears. This term is not 
invariant under the scale transformation and the subtraction of 
the $\ln\epsilon$ term breaks the invariance. The variation of 
$\ln\epsilon$ term under the scale transformation 
is finite when $\epsilon\rightarrow 0$ and should be canceled 
by the variation of the finite term (which does not 
depend on $\epsilon$) in the action since the original action 
(\ref{i}) is invariant under the scale transformation. 
Therefore  $\ln\epsilon$ term $S_{\rm ln}$ gives the Weyl 
anomaly $T$ of the action renormalized by the subtraction of 
the terms which diverge when $\epsilon\rightarrow 0$ ($d=4$) 
\be
\label{vib}
S_{\rm ln}=-{1 \over 2}
\int d^4x \sqrt{-g }T\ .
\ee
Details of the evaluation of holographic conformal anomaly 
which depends on dilaton and bulk potential have been presented 
in refs.\cite{SN,LCA}. Note that such conformal anomaly is 
evaluated on the boundary, i.e. in UV limit (asymptotically 
AdS space). Directly, this method cannot be applied for situations 
where 4d metric and dilaton on the boundary depend on the fifth, 
radial coordinate. The calculation of such 4d trace anomaly 
with the dependence on the radial coordinate which plays the role 
of scale in dual QFT has been presented in refs.\cite{BGM,LW}. 
A convenient way to calculate such radial dependent, 4d trace 
anomaly is to use the reduction of second order field equations 
to first order (see refs.\cite{BC} for explicit examples). It 
is also interesting that for Poincar\'e invariant solutions (i.e. 
when dilaton on the boundary depends only on the radial coordinate 
and 4d space is conformally flat with conformal factor depending 
only on fifth radial coordinate) the conformal anomaly is not zero. 
It represents a kind of RG improved cosmological constant. However, 
it can be shown that in UV limit (asymptotically AdS space) which 
corresponds to our situation such conformal anomaly goes to zero 
as it should be in QFT in flat background.

In this letter, we consider the maximally SUSY gauged supergravity 
where scalars parametrise a submanifold of the full scalar coset. 
As a result the bulk potential cannot be chosen arbitrarily.  
Hence, we consider the case that includes $N$ scalars and the 
coefficients $X=-{1 \over 2},~Y=0$. 
The bosonic sector of the action in this case is
\bea
\label{mul}
S={1 \over 16\pi G}\int_{M_{d+1}} d^{d+1}x \sqrt{-\hat G}
\left\{ \hat R - \sum_{\alpha=1 }^{N} {1 \over 2 }
(\hat\nabla\phi_{\alpha})^2
+\Phi (\phi_{1},\cdots ,\phi_{N} )+4 \lambda ^2 \right\}.&&
\eea
The equations of motion are given by the variation of (\ref{mul})
with respect to $\phi_{\alpha}$ and $G^{\mu \nu}$ as
\bea
\label{eqm1}
0&=&-\sqrt{-\hat{G}}{\partial \Phi(\phi_{1},\cdots ,\phi_{N})
 \over \partial \phi_{\beta}} - \partial_{\mu }\left(\sqrt{-\hat{G}}
\hat{G}^{\mu \nu}\partial_{\nu} \phi_{\beta} \right)\\
\label{eqm2}
0 &=& {1 \over d-1}\hat{G}_{\mu\nu}\left(
\Phi(\phi)+{d(d-1) \over l^2}\right)+\hat{R}_{\mu \nu}-
\sum_{\alpha=1 }^{N}{1 \over 2}
\partial_{\mu}\phi_{\alpha }\partial_{\nu}\phi_{\alpha } .
\eea
One expands $\phi_{\alpha}$ with respect to $\rho$ in the 
same way as in (\ref{viib}).
\be
g_{ij}=g_{(0)ij}+\rho g_{(1)ij}+\rho^2 g_{(2)ij}+\cdots \ ,\quad
\phi_{\alpha}=\phi_{(0)\alpha}+\rho \phi_{(1)\alpha}
+\rho^2 \phi_{(2)\alpha}+\cdots \ .
\ee
$\Phi(\phi_{1},\cdots ,\phi_{N})$ is also expanded  
\bea
\Phi&=&\Phi(\phi_{(0)})+\rho\sum_{\alpha=1}^{N}
{\partial \Phi(\phi_{(0)})
\over \partial \phi_{\alpha} }\phi_{(1)\alpha} 
+ \rho ^2\left\{ \sum_{\alpha=1}^{N} {\partial 
\Phi(\phi_{(0)}) \over
 \partial \phi_{\alpha } }\phi_{(2)\alpha } \right. \nn
&& \left. +{1 \over 2}\sum_{\alpha,\beta=1}^{N}
{\partial ^2 \Phi(\phi_{(0)})
\over \partial \phi_{\alpha}\partial \phi_{\beta} }\phi_{(1)\alpha}
\phi_{(1)\beta } \right\} + \cdots
\eea
where $\Phi(\phi_{(0)})=\Phi(\phi_{(0)1},\cdots ,\phi_{(0)N})$.

We are interested in the maximally SUSY supergravities in 
$D=d+1=3,5$ which contain $N=128,42$ scalars respectively 
(the construction of such d5 gauged supergravity has been 
given in refs.\cite{Peter}). The maximal supergravity parameterizes 
the coset $E_{11-D}/K$, where $E_{n}$ is the maximally non-compact 
form of the exceptional group $E_{n}$, and $K$ is its maximal 
compact subgroup. The $SL(N,R)$, the subgroup of $E_{n}$, can 
be parameterized via coset $SL(N,R)/SO(N)$, and we use the 
local $SO(N)$ transformations in order to diagonalize the 
scalar potential $\Phi(\phi )$ as in \cite{CGLP,MTR}
\bea
V=\Phi+4\lambda^2={d(d-1)\over N(N-2)}\left((\sum_{i=1}^{N}X_{i})^{2}-
2(\sum_{i=1}^{N}X_{i}^{2} ) \right) .
\eea
Let us briefly describe the parametrization leading to the 
action of form (6) given in ref.\cite{CGLP}. Above gauged 
supergravity case means that in $D=4,5$ we should take $N=8,6$ 
respectively. $N$ scalars $X_{i}$ which are constrained by 
\bea
\prod_{i=1}^{N}X_{i}=1
\eea
can be parameterised in terms of $(N-1)$ independent
dilatonic scalars $\phi_{\alpha}$ as follows
\bea
X_{i}=e^{-{1\over 2}b^{\alpha}_{i}\phi_{\alpha}}
\eea
Here $b_{i}^{\alpha}$ are the weight vectors of the fundamental 
representation of $SL(N,R)$, which satisfy
\bea
b_{i}^{\alpha}b_{j}^{\alpha}=8 \delta_{ij} -{8 \over N},\quad 
\sum_{i}b_{i}^{\alpha} =0 .
\eea
Then the potential has minimum at $X_{i}=1$ ($N>5$) at the point 
$\phi_{\alpha}=0$ and $V=d(d-1)$. The second derivatives of the 
potential at this minimum are given by
\bea
{\partial ^2 \Phi(\phi_{(0)})
\over \partial \phi_{\alpha}\partial \phi_{\beta} }={d(d-1)\over
N(N-2)}b_{i}^{\alpha}b_{i}^{\beta}
\eea
Here 
\bea
b_{i}^{\alpha}b_{i}^{\beta}=4(N-4)\delta^{\alpha \beta},
\eea
For maximally SUSY gauged supergravity cases described above 
(i.e. in $D=4,5$ we take $N=8,6$ respectively), we can get 
the second derivatives of the potential as
\bea
\label{sco}
{\partial ^2 \Phi(\phi_{(0)})
\over \partial \phi_{\alpha}\partial \phi_{\beta} }=2(d-2)
\delta^{\alpha \beta}.
\eea
The first derivatives of the potential are restricted 
by the leading order term in the equations of motion (\ref{eqm1}) 
\bea
\label{fco}
{\partial \Phi(\phi_{(0)}) \over \partial \phi_{\alpha} } =0 .
\eea 
We will use (\ref{sco}), (\ref{fco}) in the calculations later, 
but we also consider the case $\Phi(\phi_{(0)})=0$ which 
corresponds to the constant cosmological term.  Then, we 
introduce the parameters $a$ and $l$ and rewrite the conditions 
(\ref{sco}), (\ref{fco}) as follows:
\bea
\label{mtr}
{\partial \Phi(\phi_{(0)}) \over \partial \phi_{\alpha} } =0 \nn
{\partial ^2 \Phi(\phi_{(0)})
\over \partial \phi_{\alpha}\partial \phi_{\beta} }
={2(d-2)a\over l^2} \delta^{\alpha \beta}.
\eea
Here $a=1$ corresponds to the condition of conformal 
boundary \cite{MTR}, and $a=0$ is the case where cosmological 
term is constant. In the following calculations, 
we will use these conditions (\ref{mtr}). 
Then, $\Phi$ is expanded in a simple form 
\bea
\Phi&=&\Phi(\phi_{(0)})+ \rho ^2{a \over 2 l^2 }
\left( \sum_{\alpha=1}^{N}
2(d-2)\phi_{(1)\alpha}^2 \right)
\eea
Making the explicit calculations, after some work one can get 
the holographic conformal anomaly. For example, for 
holographic $d=2$ anomaly one finds 
\bea
S_{\rm ln}&=& -{1 \over 16\pi G}{l\over 2}\int d^{2}x
\sqrt{-g_{(0)}}
\left\{ R_{(0)} - \sum_{\alpha}^{N}{1 \over 2}
g^{ij}_{(0)}\partial_i\phi_{(0)\alpha}
\partial_j\phi_{(0)\alpha} \right\}\nn
&&\times \left( {\Phi(\phi_{(0)}) \over 2} +{2 \over l^2}\right)
\left( \Phi(\phi_{(0)})+{2 \over l^2} \right)^{-1}.
\eea
This is conformal anomaly of dual two-dimensional QFT theory 
living on the boundary of (asymptotically) AdS space. It is 
evaluated via its three-dimensional gauged SG dual. Note that 
one can consider any parametrization of scalars 
in gauged three-dimensional supergravity subject to the form 
of action (6). The dependence of the anomaly on the bulk scalar 
potential is remarkable.

In four-dimensional case the calculation of trace anomaly is 
more involved. The logarithmic term may be found as
\bea
\label{ano}
S_{\rm ln}&=&{1 \over 16\pi G}\int d^4x \sqrt{-g_{(0)}}\left[
-{1 \over 2l}g_{(0)}^{ij}g_{(0)}^{kl}\left(g_{(1)ij}g_{(1)kl}
 -g_{(1)ik}g_{(1)jl}\right) \right. \nn
&& +{l \over 2} \left(R_{(0)}^{ij}-{1 \over
2}g_{(0)}^{ij}R_{(0)}\right)g_{(1)ij} \nn
&& +{1 \over l}\sum_{\alpha}^{N}\phi_{(1)\alpha}^2
-l \sum_{\alpha}^{N}{1\over 2}\phi_{(1)\alpha }
{1 \over \sqrt{-g_{(0)}}}
\partial_i\left(\sqrt{-g_{(0)}}g_{(0)}^{ij}
\partial_j\phi_{(0)\alpha} \right) \nn
&&  -{l \over 4}\sum_{\alpha}^{N} 
\left( g_{(0)}^{ik}g_{(0)}^{jl}
g_{(1)kl}-{1 \over 2}g_{(0)}^{kl}
g_{(1)kl}g_{(0)}^{ij}\right)  \partial_i\phi_{(0)\alpha}
\partial_j\phi_{(0)\alpha}  \\
&& - {l \over 2}\left({1 \over 2}g_{(0)}^{ij}g_{(2)ij}
 -{1 \over 4}g_{(0)}^{ij}g_{(0)}^{kl}g_{(1)ik}g_{(1)jl}
+{1 \over 8}(g_{(0)}^{ij}g_{(1)ij})^2 \right)\Phi(\phi_{(0)}) \nn
&&  \left.-{a \over  l}\sum_{\alpha}^{N}\phi_{(1)\alpha }^2 
\right] \ .\nonumber
\eea 
The conditions (\ref{mtr}) are used here.
The equation of motion (\ref{eqm2}) enables one 
to express $g_{(1)ij}$
in terms of $g_{(0)ij}$ in the same way as in \cite{LCA}
\bea
\label{vibb}
g_{(1)ij}&=&\left[-R_{(0)ij}+\sum_{\alpha}^{N} {1\over 2}
\partial_i\phi_{(0)\alpha}\partial_j\phi_{(0)\alpha}\right. \nn
&& +\left. {g_{(0)ij} \over l^2}\left\{ R_{(0)}
-\sum_{\alpha}^{N} {1 \over 2}g^{ij}_{(0)}
\partial_i\phi_{(0)\alpha}\partial_j\phi_{(0)\alpha} \right\}
\times \left( {1 \over 3}\Phi(\phi_{(0)})
+{6 \over l^2} \right)^{-1} \right] \nn
&& \times \left( {1 \over 3}\Phi(\phi_{(0)})
+{2 \over l^2} \right)^{-1} \ .
\eea
In the equation (\ref{eqm1}), the terms proportional to $\rho^{-2}$
determine $\phi_{(1)}$ as follows:
\bea
\label{vii4d}
\phi_{(1)\beta}=
 -{l^{2}  \over 4(a-1)}{\partial_i \over 
 \sqrt{-g_{(0)}}}\left(\sqrt{-g_{(0)}}
g_{(0)}^{ij}\partial_j\phi_{(0)\beta } \right) . 
\eea
In the equation (\ref{eqm2}), the terms proportional to
$\rho^1$ with $\mu ,\nu =i,j$ lead to $g_{(2)ij}$
\bea
\label{viii}
g_{(2)ij}&=& \left[ -g_{(0)ij}{2a \over 3}
\sum_{\alpha}^{N}\phi_{(1)\alpha}^{2} 
-{2 \over l^2}g^{kl}_{(0)}g_{(1)ki}g_{(1)lj}
+{1\over l^2}g^{km}_{(0)}g^{nl}_{(0)}g_{(1)mn}
g_{(1)kl}g_{(0)ij} \right.\nn
&& -{2 \over l^2}g_{(0)ij}\left( {1 \over 3}\Phi(\phi_{(0)})
+{8 \over l^2} \right)^{-1}\times \left\{ {2 \over l^2}
g^{mn}_{(0)}g^{kl}_{(0)}g_{(1)km}g_{(1)ln} 
-{8a \over 3}\sum_{\alpha}^{N}\phi_{(1)\alpha}^{2} \right.\nn
&& \left.+\sum_{\alpha}^{N}g^{kl}_{(0)}
\partial_k\phi_{(1)\alpha}\partial_l\phi_{(0)\alpha}  \right\} 
\left.+\sum_{\alpha}^{N}
\partial_i\phi_{(1)\alpha}\partial_j\phi_{(0)\alpha} \right] 
\times \left( {1 \over 3}\Phi(\phi_{(0)}) \right)^{-1}.\nn
\eea
Therefore the anomaly term (\ref{ano}) is evaluated as  
\bea
\label{ano2}
S_{ln}&=&-{1 \over 2}\int d^{4}x \sqrt{-g}T, \nn
T &=&-{1 \over 8 \pi G}\left[ h_{1}R^2
+h_{2}R^{ij}R_{ij}+h_{3}R^{ij}\sum_{\alpha}^{N}
\partial_{i}\phi_{(0)\alpha}
\partial_{j}\phi_{(0)\alpha} \right. \nn
&&+h_{4}R\sum_{\alpha}^{N}g^{ij}_{(0)}\partial_{i}\phi_{(0)\alpha}
\partial_{j}\phi_{(0)\alpha}
+h_{5}\left(\sum_{\alpha}^{N}g^{ij}_{(0)}\partial_{i}\phi_{(0)\alpha}
\partial_{j}\phi_{(0)\alpha} \right)^{2} \\
&& \left.+h_{6}\sum_{\alpha}^{N}\sum_{\beta}^{N}
\left(g^{ij}_{(0)}\partial_{i}\phi_{(0)
\alpha}\partial_{j}\phi_{(0)\beta}\right)^{2}
+h_{7}\sum_{\alpha}^{N}\left({\partial_{i} \over \sqrt{-g}}
\left( \sqrt{-g}g^{ij}_{(0)}\partial_{j}
\phi_{(0)\alpha} \right) \right)^{2}\right]. \nonumber
\eea  
\nn
Here $h_{1}$, $h_{2}$, $\cdots$,  $h_{7}$ are 
\bea
\label{h1}
h_{1}&=&-h_{4}=4h_{5} =\frac{3\ (62208+22464\ \Phi+2196\ {\Phi^2}
+72\ {\Phi^3}+{\Phi^4})l^{3}}{16\ {{(6+\Phi)}^2}\ {{(18+\Phi)}^2}
\ (24+\Phi)} \\
\label{h2}
h_{2}&=& -h_{3}=4h_{6}=
-\frac{3\ (288+72\ \Phi+{\Phi^2})l^{3}}{8\ {{(6+\Phi)}^2}
\ (24+\Phi)} \nn
\label{h7}
h_7&=&\frac{((a-1)(\Phi+24)-\Phi (a-3))l^{3}}
{16 (a-1)^2 (\Phi+24)}\ .
\eea
Hereafter, we denote $\Phi(\phi_{(0)})$ by $\Phi$ and do not 
write the index $(0)$ for the simplicity. We also take $\Phi 
\to l^{2}\Phi $ as dimensionless, then we can see the dimension 
of $h$ easily, i.e. dim $h =l^3$. Thus, we found the holographic 
conformal anomaly for QFT dual from d5 gauged supergravity 
with some number of scalars which parametrises the full scalar 
coset. Note that the bulk scalar potential is an arbitrary. 
The only requirement is the form of action (\ref{mul}). One can use 
the explicit parametrization of ref.\cite{CGLP} described above 
or any other parametrization of d5 gauged 
supergravity leading to the action of the form (\ref{mul}).

Let us compare now the above conformal anomaly with the already 
known result for a single scalar. First of all, let us check the 
condition that the gravitational terms of anomaly (\ref{ano2}) 
can be written as a sum of the Gauss-Bonnet invariant $G$ 
and the square of the Weyl tensor, $F$. They are 
\bea
G &=& R^2 -4R_{ij}R^{ij}+R_{ijkl}R^{ijkl} \\
F &=& {1\over 3}R^2 -2R_{ij}R^{ij}+R_{ijkl}R^{ijkl}.
\eea
Then $R^2$ and $R_{ij}R^{ij}$ are given by
\bea
R^2&=&3G - 6F + 3R_{ijkl}R^{ijkl} \nn
R_{ij}R^{ij}&=&{1 \over 2}G - {3 \over 2}F + R_{ijkl}R^{ijkl}.
\eea
If one can rewrite the anomaly (\ref{ano2}) as 
a sum of $G$ and $F$, then $h_{1}$ and $h_{2}$ 
satisfy $3h_1+h_2=0$.  This leads to the following 
conditon for $\Phi$  
\be
3h_1 + h_2 = {3\Phi^2 (180 + \Phi^2) l^{3} 
\over 16 (6 + \Phi)^2 (18 + \Phi)^2 (24 + \Phi)} =0,
\ee 
The only solution is $\Phi =0$, i.e. constant bulk potential.
In the limit of $\Phi\to 0$, we obtain
\bea
h_{1}&\to & \frac{3\cdot 62208 l^{3}}{16\cdot 6^2 \cdot 18 ^2 \cdot
24} = { l^{3}\over 24}\nn
h_{2}&\to & -\frac{3\cdot 288  l^{3}}
{8\cdot 6^2 \cdot 24}=-{ l^{3}\over 8},
\eea
and
\bea
h_{3}&\to & +{ l^{3}\over 8},\quad h_{4}\to  -{ l^{3} \over 24}\nn
h_{5}+h_{6} &\to &-{ l^{3}\over 48}
\eea
If we take the coefficient $X=-{1\over 2},Y=0$ in Eq.(\ref{i}), 
i.e. $V=-{1 \over 2}$, in \cite{LCA}, $h_{3}$, $h_{4}$, 
$h_{5}+h_{6}$ 
agree with the single scalar case \cite{SN,LCA} exactly. 
In  this limit one gets $h_{7}$ as 
\bea
h_{7}&\to &-{ l^{3}\over 16  } ~~(a=0) \nn
h_{7}&\to &- \infty \cdot l^{3} ~~(a=1),
\eea
Hence, we find that $a=0$ case in $h_{7}$ agrees with the result in 
\cite{LCA}. Thus, we proved that 
our trace anomaly coincides with the one for constant bulk 
potential and single scalar case, where the anomaly has the 
following form:
\bea
\label{xix}
T&=&-{l^3 \over 8\pi G} 
\left[ {1 \over 8}R_{ij}R^{ij}
-{1 \over 24}R^2 \right. \nn
&& + {1 \over 2} R^{ij}\partial_i\varphi
\partial_j\varphi - {1 \over 6} Rg^{ij}
\partial_i\varphi\partial_j\varphi  \nn
&& \left. + {1 \over 4}
\left\{{1 \over \sqrt{-g}} \partial_i\left(\sqrt{-g}
g^{ij}\partial_j\varphi \right)\right\}^2 + {1 \over 3}
\left(g^{ij}\partial_i\varphi\partial_j\varphi 
\right)^2 \right]\ .
\eea
The Weyl anomaly coming from the multiplets of ${\cal N}=4$ 
supersymmetric $U(N)$ or $SU(N)$ Yang-Mills coupled with ${\cal N}=4$ 
conformal supergravity was calculated in \cite{LT}:\footnote{See Eqs.(2.5) 
and (2.6) in \cite{LT}.}
If we choose
\be
\label{xx}
{l^3 \over 16\pi G}={2N^2 \over (4\pi)^2}\ ,
\ee
and consider the background where only gravity and the real part 
of the scalar field $\varphi$ in the ${\cal N}=4$ conformal 
supergravity multiplet are non-trivial and other fields vanish, 
Eq.(\ref{xix}) exactly reproduces the result in \cite{LT}.
Unfortunately, there are no other calculations of QFT conformal anomaly
exactly, so we cannot compare our result for holografic CA when 
bulk potential is arbitrary with QFT calculation. QFT calculations of
conformal
anomaly for different interacting theories are known only in one- or
two-loop order, but not exactly.
Super Yang-Mills theory is the only case where exact result is avaliable. 

Now one considers the case $a=1$ which corresponds to the condition 
\cite{MTR}. It may look that in this situation the conformal anomaly
contains a divergence. Let us show how to take this limit 
correctly, so that divergence does not actually appear.
For the case of $a=1$, the equation (\ref{vii4d}) becomes  
\bea
{\partial_i \over \sqrt{-g_{(0)}}} \left(\sqrt{-g_{(0)}}
g_{(0)}^{ij}\partial_j\phi_{(0)\beta } \right)=0 . 
\eea
Therefore we cannot regard $\phi_{(0)}$ as
the degree of freedom on the boundary. Instead of it, we should
regard $\phi_{(1)}$, which corresponds to $d\phi/d\rho$ on the
boundary, as the independent degree of freedom. 
This would tell that $\phi_{(0)}\sim \tilde \phi \equiv 
d\phi/ d\rho$ appears as a couping constant for some 
operator ${\cal O}$ in the form of $\int d^4x \tilde\phi 
{\cal O}$.
The divergence of 
$h_{7}$ at $a=1$ should reflect this situation since
the divergence prevents us from solving $\phi_{(1)}$ in 
terms of $\phi_{(0)}$.
That is, $\phi_{(1)}$ becomes independent degree of freedom when
$a=1$.

So then, in the case of $a=1$, the anomaly is rewritten in terms of
$\phi_{(0)},\phi_{(1)}$ as 
\bea
\label{ano4}
T &=&-{1 \over 8 \pi G}\left[ h_{1}R^2
+h_{2}R^{ij}R_{ij}+h_{3}R^{ij}
\sum_{\alpha}^{N}\partial_{i}\phi_{(0)\alpha}
\partial_{j}\phi_{(0)\alpha} \right. \nn
&&+h_{4}R\sum_{\alpha}^{N}g^{ij}_{(0)}\partial_{i}\phi_{(0)\alpha}
\partial_{j}\phi_{(0)\alpha}+h_{5}
\left(\sum_{\alpha}^{N}g^{ij}_{(0)}\partial_{i}\phi_{(0)\alpha}
\partial_{j}\phi_{(0)\alpha} \right)^{2} \\
&& +h_{6}\sum_{\alpha}^{N}\sum_{\beta}^{N}
\left(g^{ij}_{(0)}\partial_{i}\phi_{(0)\alpha}
\partial_{j}\phi_{(0)\beta}\right)^{2}\nn
&&+{h_{7}\over  l^{2}}\sum_{\alpha}^{N}
\phi_{(1)\alpha}{\partial_{i} \over \sqrt{-g}}
\left( \sqrt{-g}g^{ij}_{(0)}\partial_{j}\phi_{(0)\alpha} \right) 
\left. + {h_{8}\over  l^{4}}\sum_{\alpha}^{N}
\phi_{(1)\alpha}^{2}\right]. \nonumber
\eea   
Note that from above anomaly one can get the local surface 
counterterms in the same way as in refs.\cite{LCA,MTR}.
The coefficients $h_{1}$, $h_{2}$, $\cdots$, $h_{6}$ are the same as
for the case $a \ne 1$ in (\ref{ano2}).  $h_7$ and $h_{8}$
are given by
\bea
\label{h71}
h_7&=& {(\Phi-48) l^{3} \over 4(\Phi+24)} \\
\label{h81}
h_8&=& { 2 \Phi   l^{3} \over (\Phi+24)}. 
\eea
For the constant dilaton case, eq.(\ref{ano4}) becomes
\bea
T =-{1 \over 8 \pi G}\left[ h_{1}R^2
+h_{2}R^{ij}R_{ij}\right]
\eea
It is interesting to note that coefficientes $h_{1}$, $h_{2}$
which do not depend on number of scalars in above 
expression may play the role of c-function in UV limit in 
the same way as in ref.\cite{LCA}. From the point of view of 
AdS/CFT correspondence the exponent of scalar should correspond 
to gauge coupling constant. 
Hence, this expression represents the (exact) conformal anomaly with 
radiative corrections for dual QFT. It is evaluated from the SG side.
It is a trivial task to get the anomaly for any specific bulk 
potential.

To summerize, we found explicitly non-perturbative 
conformal anomaly from gauged SG side in the situation when 
scalars respect the conformal boundary condition. 
It corresponds to the one of dual QFT living on the boundary of 
asymptotically AdS space. Using the same technique one can 
generalize the results of this work to the cases with 
the presence of other background fields (antisymmetric tensors, 
gauge fields, ...) and calculate conformal anomaly.

\section*{Acknowledgements}
The work of S.O. has been supported in part by Japan Society 
for the Promotion of Science and that of S.D.O. by CONACyT (CP, 
ref.990356 and grant 28454E).  S.O. thanks T. Sakai for useful 
discussions in Yukawa Institute of Theoretical Physics.

\end{document}